\documentclass[%
 aip,
 amsmath,amssymb,
 reprint,%
]{revtex4-1}

\usepackage{amsmath, amsthm, amssymb}
\usepackage{graphicx}
\usepackage{dcolumn}
\usepackage{bm}
\usepackage{bbm}
\usepackage{color}
\usepackage[usenames,dvipsnames]{xcolor}
\usepackage{hyperref}
\usepackage{enumerate}
\usepackage{ulem}
\usepackage{textcomp}

\newcommand{\ket}[1]{| #1 \rangle}

\newcommand{\Tr}{\operatorname{Tr}}

\newcommand{\ba}{\begin{eqnarray}}
\newcommand{\ea}{\end{eqnarray}}

\begin{document}

\title{Fast self-testing Quantum Random Number Generator based on homodyne detection}

\author{Davide Rusca}
\affiliation{D\'epartment de Physique Appliqu\'ee, Universit\'e de Gen\`eve, 1211 Gen\`eve, Switzerland}
\author{Hamid Tebyanian}
\affiliation{Dipartimento di Ingegneria dell’Informazione, Università degli Studi di Padova, Padova, Italia}
\author{Anthony Martin}
\affiliation{D\'epartment de Physique Appliqu\'ee, Universit\'e de Gen\`eve, 1211 Gen\`eve, Switzerland}
\author{Hugo Zbinden}
\affiliation{D\'epartment de Physique Appliqu\'ee, Universit\'e de Gen\`eve, 1211 Gen\`eve, Switzerland}

\begin{abstract}

Self-testing and Semi-Device Independent protocols are becoming the preferred choice for quantum technologies, being able to certify their quantum nature with few assumptions and simple experimental implementations. In particular for Quantum Random Number Generators the possibility of monitoring in real time the entropy of the source only by measuring the input/output statistics is a characteristic that no other classical system could provide. The cost of this new possibility is not necessarily increased complexity and reduced performance. Indeed, here we show that with a simple optical setup consisting of commercially available components, a high bit generation rate can be achieved. We manage to certify 145.5~MHz of quantum random bit generation rate. 

\end{abstract}

\maketitle

\section{Introduction}

Quantum random number generators have been a developing topic in the past two decades. The interest of these devices resides in the fact that the randomness of the output string can be proven thanks to the intrinsic nature of quantum mechanics and does not need a stochastic model in order to evaluate the amount entropy the experiment can produce.

However, while there exist many examples of quantum random number generators (QRNGs) that exploit many different quantum phenomena \cite{stefanov2000,jennewein2000,gabriel2010,qi2010,abellan2014,sanguinetti2014}, the challenge resides into proving that the randomness produced has, actually, a quantum origin. To do so, the device must be completely characterized in order to separate all possible sources of classical noise that could be foreseen by a malicious party. This first class of QRNGs is often referred as Device Dependent QRNG since their behaviour is strongly related to the characterization of the device. 

A different approach is instead given by the Device Independent approach \cite{colbeckPhD,pironio2010,colbeck_private_2011}. In this case it is possible to certify the randomness of the output in the most paranoid scenario in which the device itself is built by an adversary. While this approach is interesting and gives the highest level of security for a QRNG device, it has itself some drawbacks. First of all, the experiment relies still on some assumptions that must be verified, such as the space-like separation of the two measurement sites, but the greatest roadblock toward applications is the complexity of the devices and their low throughput rate, which is orders of magnitude lower than for standard DD-QRNG.

For this reason, recently a new approach has been investigated, the Semi-Device Independent or self-testing QRNG~\cite{li2011,vallone2014,lunghi2015,canas2014,Cao2015,Marangon2017,Cao2016,Xu2016,Brask2017a,Gehring2018,Thibault2019}. 
The idea is, with few assumptions on the device, implement a QRNG as simple as the existing commercial devices with high degree of security.
Recently, a couple of different approaches have been proposed using assumptions on the dimension of the produced quantum states, on their overlap or on their energy.

In this work, we develop a new QRNG using an energy assumption featuring a simple and practical implementation based on homodyne measurement with a performance of hundreds of certified Mbits/s. Note that a similar approach based on a heterodyne measurement has been demonstrated independently~\cite{avesani2020}.

\section{Scheme}

As in a previous experiment~\cite{rusca2019} the security framework is based on recent theoretical work~\cite{himbeeck2017,himbeeck2018}. The device can be modelled as a prepare and measure scenario. The source prepares one of two quantum states in an optical signal depending on a binary input $x$ and sends them to a measurement device that outputs a binary value $b$. The prepared state and the measurement may depend on a correlated random variable $\lambda$. The probability of each output conditioned on the input value can be then written as:

\begin{equation}
\label{eq.pbx}
p(b|x) = \sum_\lambda p(\lambda) \Tr[ \rho^\lambda_x M^\lambda_b ] ,
\end{equation}
where $\rho^\lambda_x$ is the quantum state prepared by the source and  $M^\lambda_b$ is the POVM element corresponding to the output $b$ with $x,b \in \{0,1\}$. In order to certify the quantum randomness in output $b$ and separate it from classical noise represented by the classical variable $\lambda$ we make use of the analysis presented in Ref.~\cite{himbeeck2018}.

Without any constraint on the prepared state and without any additional assumptions, the measured input output probabilities (correlated classically by an unknown classical hidden variable) could be easily described by a deterministic model. In our case, the two states can be arbitrarily chosen but their energy must be limited. This assumption is referred as "average energy" assumption, since the quantity to be bounded is the following:

\begin{equation}
	\label{eq.energycons}
	\sum_{\lambda,x} p(\lambda) \Tr[\rho^\lambda_x N] \leq \omega \,,
\end{equation}
which represents the upper bound on the average energy (normalized with respect to the lowest photon energy of the optical signal) transmitted between the source and the measurement, $N$ in this formula represent the photon number operator and $\omega$ the bound chosen. The normalized energy is in this way given by the average number of photon transmitted in each signal.

Van Himbeeck et al. \cite{himbeeck2017} showed that for a fixed value $\omega$ the set of all possible quantum correlations is larger than the set of deterministic correlations. Furthermore in the following work \cite{himbeeck2018} it was proven that if an input/output distribution belongs to the former set but not to the latter then genuine randomness can be certified.
In order to quantify the entropy produced by the input/output statistics of the experiment, the authors developed a semi-definite program (SDP) that returns a lower bound on the conditional Shannon Entropy $H(B|X,\Lambda)$. This bound can be used as witness to certify the amount of genuine quantum randomness. This witness corresponds to a linear function $\gamma[p]-\zeta[\omega]$ that depends only on the input/output probabilities $p$ and the average energy bound $\omega$. The witness defined before can be then used for a semi-device independent protocol where an entropy threshold $h$ is fixed beforehand and the previously defined witness is tailored over the expected behaviour of the device. After running the experiment $n$ times, we check that the linear witness is greater than the threshold $h$:

\begin{equation}\label{eq.3}
    \gamma[f]-\zeta[\omega] \geq h,
\end{equation}
where the witness is evaluated with the experimental input/output frequencies $f$ measured from the experiment input/output results.

If the measured data passes the test in Eq.~\ref{eq.3}, the randomness contained by the output sequence is certified to be \cite{himbeeck2018}:

\begin{equation}
	\label{eq.finite_rate}
	H^{\epsilon'}_{\min}(\bm{B}|\bm{X},\bm \Lambda)\geq n \left(h - c \sqrt{\tfrac{\log(\epsilon/2)}{n}} - d \tfrac{\log(\epsilon/2)}{n}\right)\,.
\end{equation}
where $H_{\min}^{\epsilon'}(\bm{B}|\bm{X},\bm{\Lambda})$ is the worst-case conditional smooth min-entropy and can be interpreted as the amount of bits that a strong extractor can output from the raw bit sequence generated by the experiment (See Ref.\cite{rusca2019,himbeeck2018} for more details).

\begin{figure}[t]
  \centering
  \includegraphics[width=\columnwidth]{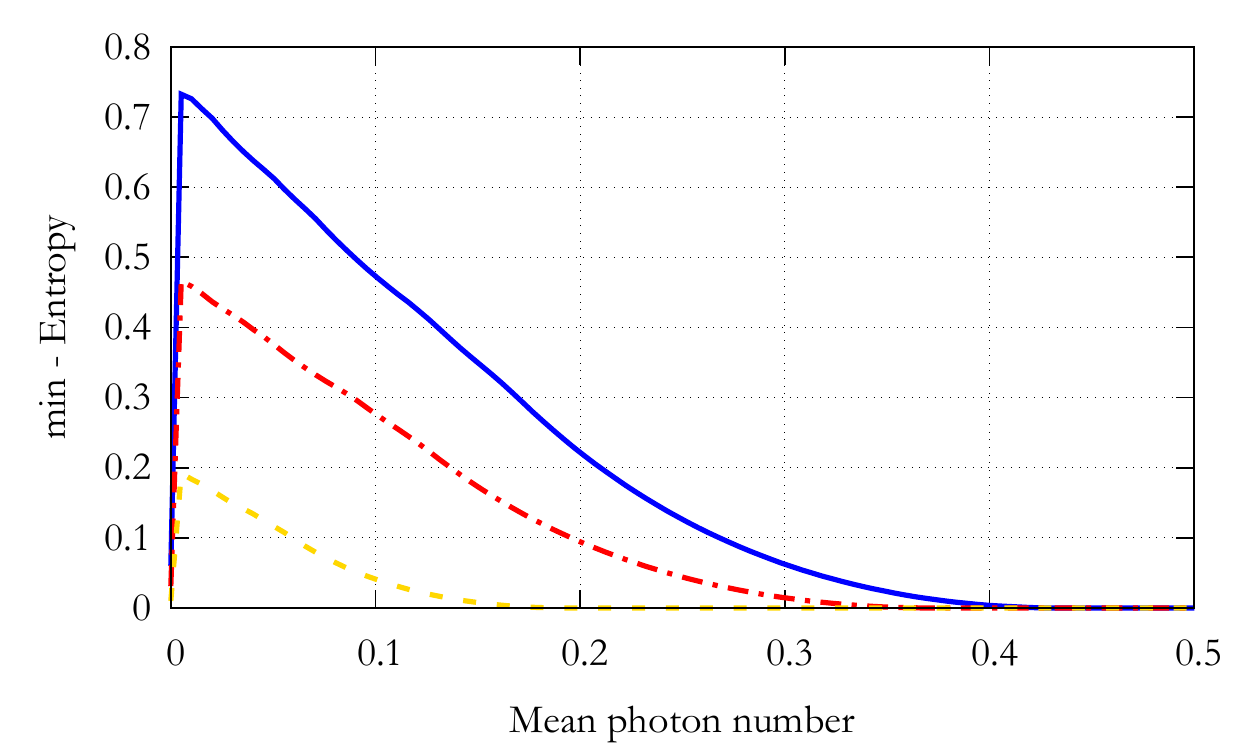}
  \caption{Extractable randomness per bit with respect to the average photon number of the input state, using the BPSK encoding scheme. The continuous line (BLUE) corresponds to a measurement strategy that reaches the Helstrom limit, the dashed-dotted line (RED) to a perfect homodyne detection scheme and the dashed line (yellow) to an homodyne detection with added white noise ($p_{noise} = 0.39$ see Eq.~\ref{eq.p_n}).}
  \label{fig.EvsM}
\end{figure}

\begin{figure}[t]
  \centering
  \includegraphics[width=\columnwidth]{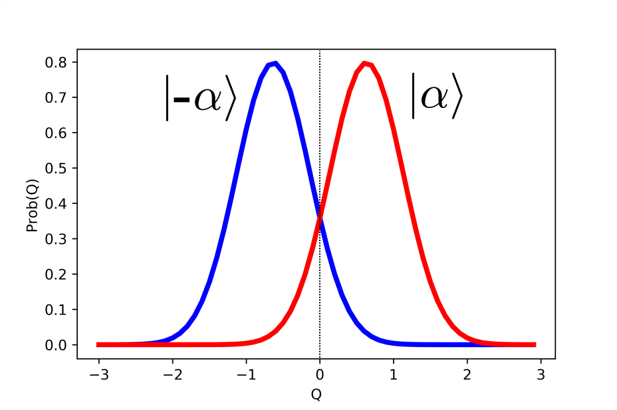}
  \caption{Quadrature distribution of two coherent states $\ket{-\alpha}$ and $\ket{\alpha}$. The maximum of the two distribution is, respectively, in the negative part of the graph and in the positive part. The best strategy to discriminate between the two state is using the sign of the quadrature.}
  \label{fig.coh}
\end{figure}

\begin{figure}[t]
  \centering
  \includegraphics[width=\columnwidth]{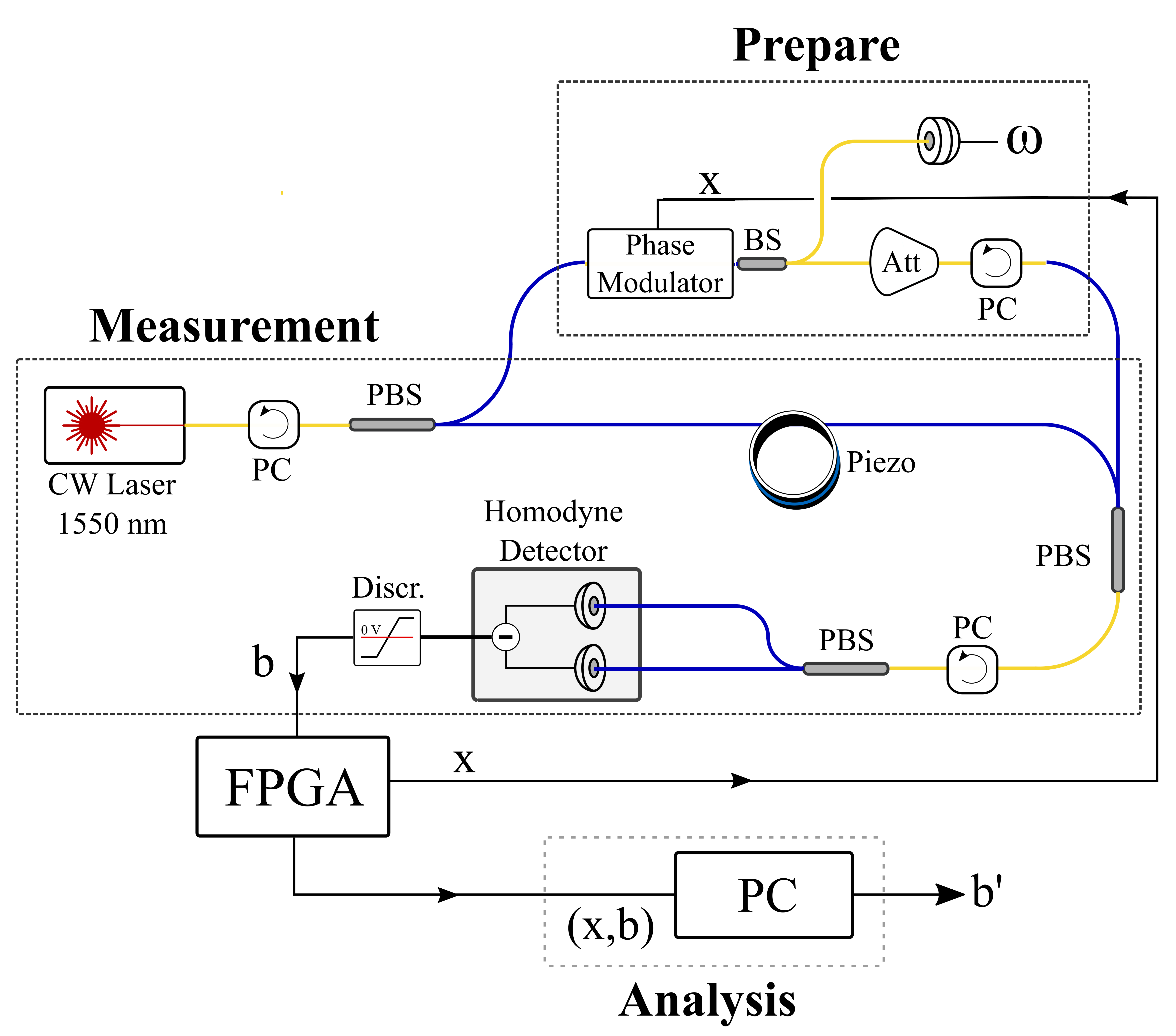}
  \caption{The experimental setup corresponds to a CW Laser at 1550 nm that injects light into a MZI. The input port is a PBS in order to modulate the amount of energy going into each arm of the interferometer. The top arm, going into the Prepare stage, correspond to the state preparation where the light is modulated by a phase modulator and the average energy is measured by a system of BS, attenuator and linear detector. The bottom arm corresponds to the Local Oscillator of the homodyne measurement, the fiber is wrapped around a piezo in order to stabilize the interferometer phase. The measurement is carried out with two balanced linear detectors.}
  \label{fig.Setup}
\end{figure}

\section{Implementation}

For the implementation we used the Binary Phase Shift- Keying (BPSK) scheme where the source prepares two coherent states with the same average photon number and a $\pi$ phase difference, i.e. $\ket{\alpha}$ and $\ket{-\alpha}$. By choosing either state with probability $1/2$ the average energy bound $\omega$ must be greater than $|\alpha|^2$. This choice is motivated by the fact that implementing such a source is easy and can allow for high repetition rates. Within, all measurement strategies we need to determine the one that can discriminate the two produced state in the best way. Fig.~\ref{fig.EvsM} shows the comparison of different strategies. The best strategy is given by the min-error discrimination measurement that allows to achieve the Helstrom limit. However, a perfect homodyne measurement, in which the two states are distinguished with respect to the sign of their quadrature (see the graphical representation in Fig.~\ref{fig.coh}), does not fall far behind, not even when noise is take into consideration. This given a practical solution even for the measurement itself.

 This scheme is implemented as shown in Fig. \ref{fig.Setup}. We generate the two desired states of light by modulating a coherent state with a phase modulator and we implement a homodyne measurement in order to project the states into the chosen quadrature. The setup is completely fibered, it is composed by a continuous-wave (CW) laser at telecommunication wavelength (1550nm) which injects light into a balanced Mach-Zehnder interferometer (MZI). The input port of the MZI consist of an optical system of polarization controller (PC) and fiber polarization beam splitter (PBS) to to adjust the power going into each arm of the MZI. As it is shown in Fig.~\ref{fig.Setup} the lower arm of the interferometer corresponds to the local oscillator of the homodyne measurement, and the upper MZI arm to the preparation stage of the experiment. In this part the states are modulated and attenuated to the energy value required by the energy assumption taken in the protocol. A set of 50/50 beam splitter, linear photo-diode and attenuator allows to monitor in real time the average mean photon number of the states produced by the preparation stage. In order to obtain this value, we divide the power measured during a run of the experiment for the number of signal exchanged during this time. The result is then multiplied by the lowest signal photon energy in the spectral band (this gives an upper bound on the mean photon number of our signal). This puts us in the worst case scenario where a duty cycle of 100\% is considered.  We would like to stress that these three components (beam splitter, attenuator and linear detector) are the only part of the experiment that must be trusted and characterized. Indeed the main assumption of the scheme can be defined as the energy of the pulses (mean photon number) going out of the preparation stage. 
The phase modulator is controlled by a binary input sent by a Field Programmable Gate Array (FPGA) at a repetition rate of 1.25~Gbits/s.

The prepared states are then recombined with the LO by a set of two PBSs and PC (that serve the purpose of a variable beam splitter) in order to balance the power transmitted to the balanced photo-diode (Thorlabs PDB480C-AC). The analog signal coming from the homodyne detector is discriminated between positive and negatives values. Which corresponds to the discrimination between positive and negative quadrature values. The binary output $b$, generated in this way, is then collected by the FPGA. Electrical delay lines are used in order to synchronize the input ($x$) and output ($b$). Moreover the discrimination is triggered by a clock signal sent by the FPGA and controlled in such a way that the discrimination windows is optimized to obtain the best discrimination. To stabilize the phase on the interferometer, a digital optimization is set-up over the correlation measured. A feedback signal is sent to a piezo-electric cylinder over which 2 m of fiber are wrapped. The stability of the interferometer is then achieved without the need of an additional source of light. Passive stability of the set-up could be achieved by shortening the arms of the interferometer (currently of 6 m each) or by integrating the scheme in a photonic circuit.

Input and output are collected by the FPGA and forwarded to an offline PC that evaluates the conditional probabilities $p(b|x)$ and that calculates the extraction rate certified by the semi-device independent protocol. 

\begin{figure}[t]
  \centering
  \includegraphics[width=\columnwidth]{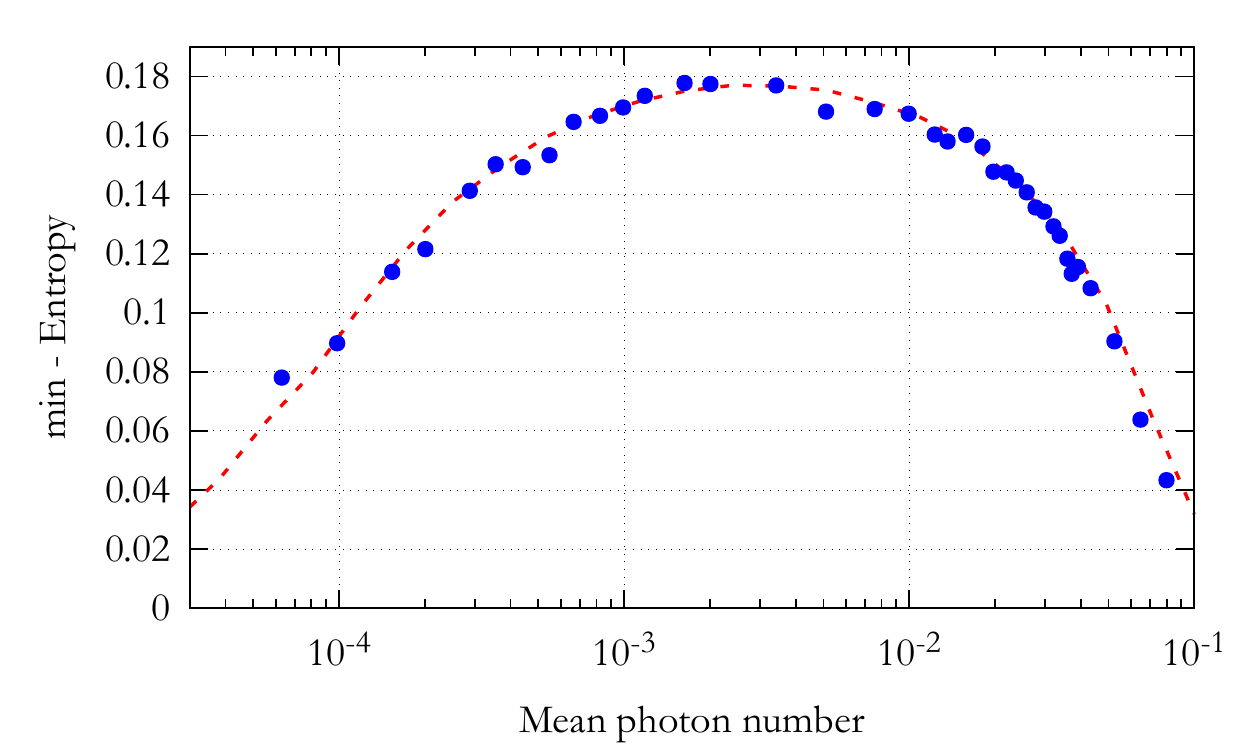}
  \caption{Amount of entropy per bits per generated pulse vs the average mean photon number. The average energy bound is chosen equal to the mean photon number to obtain the maximum achievable entropy. The dots corresponds to the experimental measured values, the dashed line corresponds to the theoretical model used to simulate the device behaviour.}
  \label{fig.Entropy}
\end{figure}

\section{Results}

First we measure the dependence of maximum extractable randomness with respect to the chosen energy bound (expressed in the mean photon number of the prepared pulses). As shown in Fig. \ref{fig.Entropy} the amount of extractable randomness has a maximum around $10^{-3}$ to $10^{-2}$ photon per pulse. This value is the result of a trade off between a small enough energy in order to obtain a uniform probability distribution, but high enough energy in order to be able to distinguish between the two input values without being dominated by electrical noise. 

\begin{figure}[t]
  \centering
  \includegraphics[width=\columnwidth]{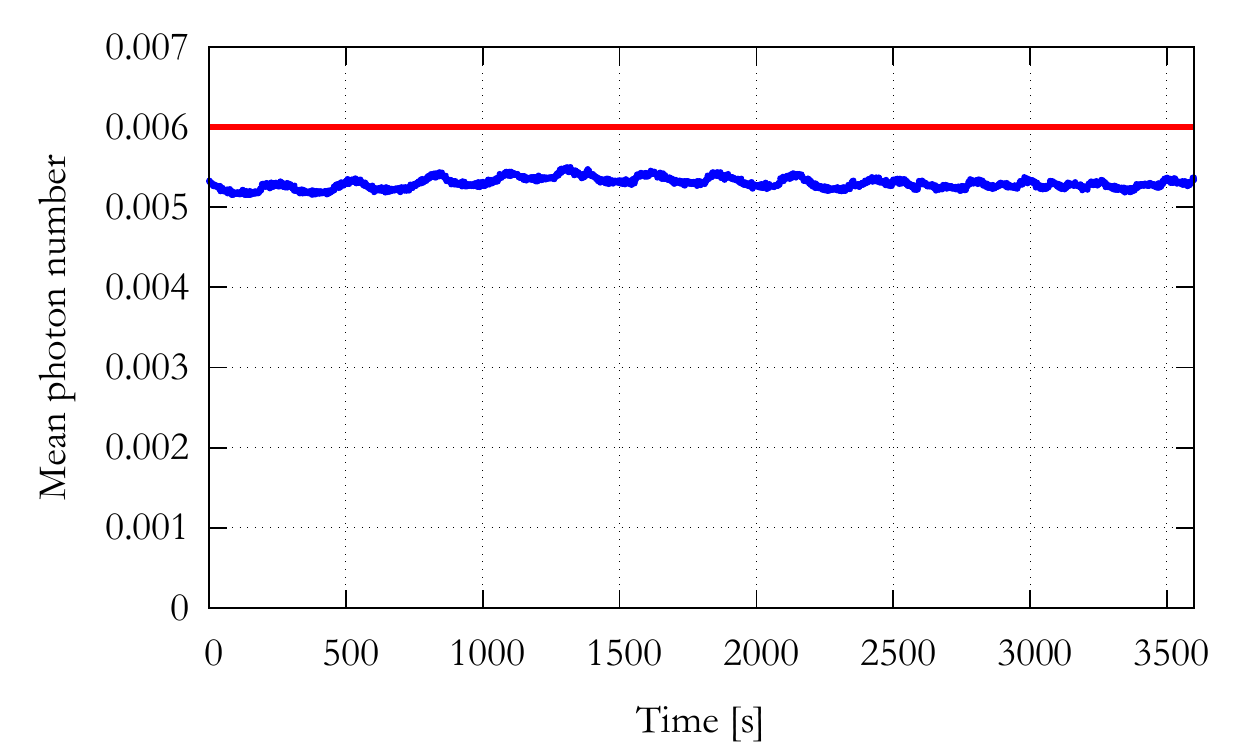}
  \caption{Measured mean photon number vs time. The points correspond to the verified mean photon number measured in the experiment. The solid line is the energy bound. As it can be seen the assumption of our experiment is never violated.}
  \label{fig.pow}
\end{figure}

\begin{figure}[t]
  \centering
  \includegraphics[width=\columnwidth]{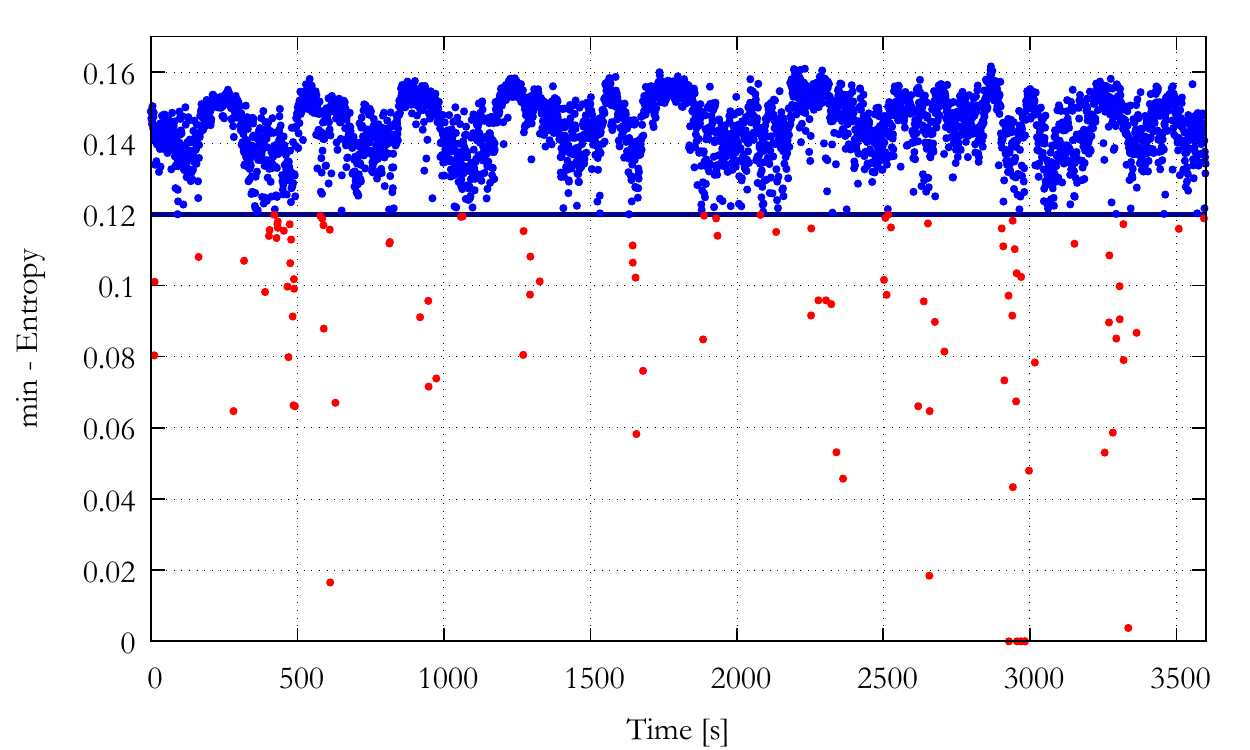}
  \caption{Amount of entropy per bits per generated pulse vs time. Each point in the figure corresponds to a 1~s measurement. The solid line corresponds to the threshold $h$ and each point above (blue) is a successful measurement whilst each point below this line (red) correspond to a failure.}
  \label{fig.stab}
\end{figure}

In order to estimate the amount of unwanted "classical" or not trusted noise in the experiment a simple model has been used to approximate the experimentally measured data points. The conditional probabilities have been modelled as follows:

\begin{equation}\label{eq.p_n}
    p(b|x) = (1-p_{noise})p^{id}(b|x) + p_{noise}\frac{1}{2},
\end{equation}
where $p^{id}(b|x)$ corresponds to the ideal homodyne measurement with no added noise and perfect state preparation and $p_{noise}$ is an arbitrary parameter used as a figure of merit in the modelling (in the experiment this value is $p_{noise} = 0.39$). This model corresponds to a system that works as expected in an ideal way with probability $(1-p_{noise})$ and with probability $p_{noise}$ it will output a values completely uncorrelated with the input. This probability represents all possible imperfection of the experiment, like state preparation flaws, electrical noise in the detection scheme, etc... However the only purpose of this value for our protocol is to be a figure of merit for the experiment, since it never appears as a parameter in the security proof. This represent the advantage of the self-testing approach, in fact with the sole analysis of the input and output statistics plus few reasonable assumption it is possible to certify the amount of entropy generated without the complete characterization of the device. In a completely device dependent scenario the probability $p_{noise}$ should be perfectly characterized and calculated a priory and then monitored through the all functioning of the experiment in order to certify the genuine randomness of the output.

Once the optimal energy bound has been estimated, we carried out a longer measurement of 1 hour in order to test the stability and resilience of the experiment. Following the semi-DI protocol presented before it was chosen a value for the threshold $h$ which represent the asymptotic extractable entropy. Each second the input/output frequencies are estimated and the assumption is verified. As it can be seen from Fig.~\ref{fig.pow} the measured power was never higher than the fixed threshold. The operating mean photon number chosen is around $5\times10^{-3}$ by optimizing the entropy per bit and by verifying that the correlation generated were sufficient for the feedback stabilization loop to work. Fig.~\ref{fig.stab} shows the entropy as a function of the time. For each point it is verified that the entropy generated is higher than the previously fixed threshold. If this condition is verified the extraction ratio is given by Eq.~\ref{eq.finite_rate} otherwise the protocol must abort. The choice of the threshold value comes with a trade-off between the amount of extractable bits per pulse generated and the amount of successful rounds in the experiment. In order to maximize the average rate of certified random bits throughout the whole experiment it was chose a threshold value of $h = 0.12$ to which corresponded a probability of successful rounds of $97\%$. This two values allow the experiment to certify a repetition rate of genuine quantum bits of 145.5~MHz.

\section{Conclusion}

The QRNG presented in this work is a simple yet performant implementation of the semi-DI protocol based on energy bounds. We achieved a random bit rate of 145.5 MHz for a measurement during 1h. The advantage of the system not only relies in its high speed, but also in the straight forward implementation, which is highly compatible with a possible integrated optics implementation.

\section*{Acknowledgements} We would like to thank Nicolas Brunner, Thomas van Himbeeck and Stefano Pironio for all the helpful discussions. We acknowledge the support of European Union’s Horizon 2020 program under the Marie Sk\l{}odowska-Curie project QCALL (GA 675662), the Swiss National Science Foundation (Bridge project “Self-testing QRNG”) and the EU Quantum Flagship project QRANGE.

\bibliography{maxavg_qrng}

\appendix

\end{document}